
%
\font\twelverm=cmr10 scaled 1200    \font\twelvei=cmmi10 scaled 1200
\font\twelvesy=cmsy10 scaled 1200   \font\twelveex=cmex10 scaled 1200
\font\twelvebf=cmbx10 scaled 1200   \font\twelvesl=cmsl10 scaled 1200
\font\twelvett=cmtt10 scaled 1200   \font\twelveit=cmti10 scaled 1200
\skewchar\twelvei='177   \skewchar\twelvesy='60

    \font\eleveni=cmmi10 scaled 1095
\font\elevensy=cmsy10 scaled 1095



\skewchar\eleveni='177   \skewchar\elevensy='60

\newfam\mibfam%

\def\twelvepoint{\normalbaselineskip=12.4pt
  \abovedisplayskip 12.4pt plus 3pt minus 9pt
  \belowdisplayskip 12.4pt plus 3pt minus 9pt
  \abovedisplayshortskip 0pt plus 3pt
  \belowdisplayshortskip 7.2pt plus 3pt minus 4pt
  \smallskipamount=3.6pt plus1.2pt minus1.2pt
  \medskipamount=7.2pt plus2.4pt minus2.4pt
  \bigskipamount=14.4pt plus4.8pt minus4.8pt
  \def\rm{\fam0\twelverm}          \def\it{\fam\itfam\twelveit}%
  \def\sl{\fam\slfam\twelvesl}     \def\bf{\fam\bffam\twelvebf}%
  \def\mit{\fam 1}                 \def\cal{\fam 2}%
  \def\tt{\twelvett}
  \textfont0=\twelverm   \scriptfont0=\tenrm   \scriptscriptfont0=\sevenrm
  \textfont1=\twelvei    \scriptfont1=\teni    \scriptscriptfont1=\seveni
  \textfont2=\twelvesy   \scriptfont2=\tensy   \scriptscriptfont2=\sevensy
  \textfont3=\twelveex \scriptfont3=\twelveex \scriptscriptfont3=\twelveex
  \textfont\itfam=\twelveit
  \textfont\slfam=\twelvesl
  \textfont\bffam=\twelvebf \scriptfont\bffam=\tenbf
  \scriptscriptfont\bffam=\sevenbf
  \normalbaselines\rm}




\def\beginlinemode{\endmode
  \begingroup\parskip=0pt \obeylines\def\\{\par}\def\endmode{\par\endgroup}}
\def\beginparmode{\endmode
  \begingroup \def\endmode{\par\endgroup}}
\let\endmode=\par
{\obeylines\gdef\
{}}
\def\singlespace{\baselineskip=\normalbaselineskip}
\def\oneandathirdspace{\baselineskip=\normalbaselineskip
  \multiply\baselineskip by 4 \divide\baselineskip by 3}
\def\oneandahalfspace{\baselineskip=\normalbaselineskip
  \multiply\baselineskip by 3 \divide\baselineskip by 2}
\def\doublespace{\baselineskip=\normalbaselineskip \multiply\baselineskip by 2}

\newcount\firstpageno
\firstpageno=2
\footline={\ifnum\pageno<\firstpageno{\hfil}\else{\hfil\twelverm\folio\hfil}\fi}
\def\toppageno{\global\footline={\hfil}\global\headline
  ={\ifnum\pageno<\firstpageno{\hfil}\else{\hfil\twelverm\folio\hfil}\fi}}
\let\rawfootnote=\footnote              
\def\footnote#1#2{{\rm\singlespace\parindent=0pt\parskip=0pt
  \rawfootnote{#1}{#2\hfill\vrule height 0pt depth 6pt width 0pt}}}
\def\raggedcenter{\leftskip=4em plus 12em \rightskip=\leftskip
  \parindent=0pt \parfillskip=0pt \spaceskip=.3333em \xspaceskip=.5em
  \pretolerance=9999 \tolerance=9999
  \hyphenpenalty=9999 \exhyphenpenalty=9999 }
\def\dateline{\rightline{\ifcase\month\or
  January\or February\or March\or April\or May\or June\or
  July\or August\or September\or October\or November\or December\fi
  \space\number\year}}
\def\received{\vskip 3pt plus 0.2fill
 \centerline{\sl (Received\space\ifcase\month\or
  January\or February\or March\or April\or May\or June\or
  July\or August\or September\or October\or November\or December\fi
  \qquad, \number\year)}}


\hsize=6.5truein
\hoffset=0.05truein    
\vsize=8.9truein
\voffset=0.05truein   
\parskip=\medskipamount
\def\\{\cr}
\twelvepoint            
\doublespace            
\overfullrule=0pt       





\def\title                      
  {\null\vskip 3pt plus 0.1fill
   \beginlinemode \doublespace \raggedcenter \bf}

\def\author                     
  {\vskip 3pt plus 0.25fill \beginlinemode
   \singlespace \raggedcenter}

\def\affil                      
  {\vskip 3pt plus 0.1fill \beginlinemode
   \oneandathirdspace \raggedcenter \sl}

\def\abstract                   
  {\vskip 3pt plus 0.2fill \beginparmode
   \oneandathirdspace ABSTRACT: }

\def\resume                   
  {\vskip 3pt plus 0.2fill \beginparmode
   \oneandahalfspace RESUME: }

\def\endtopmatter               
  {\endpage                     
   \body}

\def\body                       
  {\beginparmode}               

\def\head#1{                    
  \goodbreak\vskip 0.5truein    
  {\immediate\write16{#1}
   \raggedcenter \uppercase{#1}\par}
   \nobreak\vskip 0.25truein\nobreak}

\def\beneathrel#1\under#2{\mathrel{\mathop{#2}\limits_{#1}}}

\def\refto#1{$^{#1}$}    

\def\references         
  {\head{References}      
   \beginparmode
   \frenchspacing \parindent=0pt \leftskip=1truecm
   \parskip=8pt plus 3pt \everypar{\hangindent=\parindent}}

\def\bibliog         
  {      
   \beginparmode
   \frenchspacing \parindent=0pt \leftskip=1truecm
   \parskip=8pt plus 3pt \everypar{\hangindent=\parindent}}

\gdef\refis#1{\item{#1.\ }}                     

\gdef\journal#1, #2, #3, 1#4#5#6{       
    {\sl #1~}{\bf #2}, #3 (1#4#5#6)}        

\def\pr{\journal Phys. Rev., }

\def\prb{\journal Phys. Rev. B, }

\def\prl{\journal Phys. Rev. Lett., }

\def\epl{\journal Euro. Phys. Lett.,}

\def\rmp{\journal Rev. Mod. Phys., }

\def\pl{\journal Phys. Lett., }

\def\jpc{\journal J. Phys. C, }

\def\jetp{\journal Sov. Phys. JETP, }

\def\jetl{\journal Sov. Phys. JETP Letters, }

\def\zpb{\journal Zeit. Phys. B., }

\def\jdp{\journal J.  Phys. (Paris), }

\def\jdc{\journal J.  Phys. (Paris) Colloq., }

\def\jpjap{\journal J. Phys. Soc. Japan, }

\def\physica{\journal Physica B, }

\def\phl{\journal Phil. Mag.,}

\def\jmmm{\journal J. Mag. Mat., }

\def\endreferences{\body}

\def\figurecaptions             
  {\endpage
   \beginparmode
   \head{Figure Captions}
}

\def\endfigurecaptions{\body}

\def\tablecaptions             
  {\endpage
   \beginparmode
   \head{Table Captions}
}

\def\endpage                    
  {\vfill\eject}

\def\endpaper                   
  {\endmode\vfill\supereject}

\def\endit
  {\endpaper\end}


\def\heading                            
  {\vskip 0.5truein plus 0.1truein      
   \beginparmode \def\\{\par} \parskip=0pt \singlespace \raggedcenter}

\def\subheading                         
  {\vskip 0.25truein plus 0.1truein     
   \beginlinemode \singlespace \parskip=0pt \def\\{\par}\raggedcenter}

\def\tag#1$${\eqno(#1)$$}

\def\align#1$${\eqalign{#1}$$}

\def\aligntag#1$${\gdef\tag##1\\{&(##1)\cr}\eqalignno{#1\\}$$
  \gdef\tag##1$${\eqno(##1)$$}}

\def\endaligntag{}

\def\overset#1\to#2{{\mathop{#2}^{#1}}}
\def\underset#1\to#2{{\mathop{#2}_{#1}}}


\def\ref#1{Ref.~#1}                     
\def\Ref#1{Ref.~#1}                     
\def\[#1]{[\cite{#1}]}
\def\cite#1{{#1}}
\def\(#1){(\call{#1})}
\def\call#1{{#1}}
\def\taghead#1{}
\def\frac#1#2{{#1 \over #2}}

\def\12{{1\over2}}

\def\sla{\raise.15ex\hbox{$/$}\kern-.57em}
\def\leaderfill{\leaders\hbox to 1em{\hss.\hss}\hfill}
\def\twiddle{\lower.9ex\rlap{$\kern-.1em\scriptstyle\sim$}}
\def\bigtwiddle{\lower1.ex\rlap{$\sim$}}
\def\gtwid{\mathrel{\raise.3ex\hbox{$>$\kern-.75em\lower1ex\hbox{$\sim$}}}}
\def\ltwid{\mathrel{\raise.3ex\hbox{$<$\kern-.75em\lower1ex\hbox{$\sim$}}}}
\def\square{\kern1pt\vbox{
\hrule height 1.2pt\hbox{\vrule width 1.2pt\hskip 3pt
\vbox{\vskip 6pt}\hskip 3pt\vrule width 0.6pt}\hrule height 0.6pt}\kern1pt}
\def\tdot#1{\mathord{\mathop{#1}\limits^{\kern2pt\ldots}}}

\def\pmb#1{\setbox0=\hbox{#1}%
  \kern-.025em\copy0\kern-\wd0
  \kern  .05em\copy0\kern-\wd0
  \kern-.025em\raise.0433em\box0 }

\def\refto#1{$^{#1}$}

\def\eps{\epsilon}
\def\3he{{$^3${\rm He}}}


\def\slD{\raise.15ex\hbox{$/$}\kern-.57em\hbox{$D$}}
\def\dsl{\raise.15ex\hbox{$/$}\kern-.57em\hbox{$\Delta$}}
\def\slp{{\raise.15ex\hbox{$/$}\kern-.57em\hbox{$\partial$}}}
\def\nsl{\raise.15ex\hbox{$/$}\kern-.57em\hbox{$\nabla$}}
\def\sla{\raise.15ex\hbox{$/$}\kern-.57em\hbox{$\rightarrow$}}
\def\slla{\raise.15ex\hbox{$/$}\kern-.57em\hbox{$\lambda$}}
\def\gtwid{\raise.3ex\hbox{$>$\kern-.75em\lower1ex\hbox{$\sim$}}}
\def\ltwid{\raise.3ex\hbox{$<$\kern-.75em\lower1ex\hbox{$\sim$}}}

\def\12{{1\over2}}
\def\a{\alpha}
\def\part{\partial}
\def\la{\lambda}

\def\be{{\beta}}

\def\bethlogo{\vbox{\bf \line{\hrulefill}
    \kern-.5\baselineskip
    \line{\hrulefill\phantom{ ELIZABETH A. MASON }\hrulefill}
    \kern-.5\baselineskip
    \line{\hrulefill\hbox{ ELIZABETH A. MASON }\hrulefill}
    \kern-.5\baselineskip
    \line{\hrulefill\phantom{ 1411 Chino Street }\hrulefill}
    \kern-.5\baselineskip
    \line{\hrulefill\hbox{ 1411 Chino Street }\hrulefill}
    \kern-.5\baselineskip
    \line{\hrulefill\phantom{ Santa Barbara, CA 93101 }\hrulefill}
    \kern-.5\baselineskip
    \line{\hrulefill\hbox{ Santa Barbara, CA 93101 }\hrulefill}
    \kern-.5\baselineskip
    \line{\hrulefill\phantom{ (805) 962-2739 }\hrulefill}
    \kern-.5\baselineskip
    \line{\hrulefill\hbox{ (805) 962-2739 }\hrulefill}}}
\def\lisalogo{\vbox{\bf \line{\hrulefill}
    \kern-.5\baselineskip
    \line{\hrulefill\phantom{ LISA R. GOODFRIEND }\hrulefill}
    \kern-.5\baselineskip
    \line{\hrulefill\hbox{ LISA R. GOODFRIEND }\hrulefill}
    \kern-.5\baselineskip
    \line{\hrulefill\phantom{ 6646 Pasado }\hrulefill}
    \kern-.5\baselineskip
    \line{\hrulefill\hbox{ 6646 Pasado }\hrulefill}
    \kern-.5\baselineskip
    \line{\hrulefill\phantom{ Santa Barbara, CA 93108 }\hrulefill}
    \kern-.5\baselineskip
    \line{\hrulefill\hbox{ Santa Barbara, CA 93108 }\hrulefill}
    \kern-.5\baselineskip
    \line{\hrulefill\phantom{ (805) 962-2739 }\hrulefill}
    \kern-.5\baselineskip
    \line{\hrulefill\hbox{ (805) 962-2739 }\hrulefill}}}

\def\la{{\lambda}}

\def\low#1{\lower.5ex\hbox{${}_#1$}}
\def\ltwid{\raise.3ex\hbox{$<$\kern-.75em\lower1ex\hbox{$\sim$}}}

\def\om{{\omega}}

\def\psl{\raise.15ex\hbox{$/$}\kern-.57em\hbox{$\partial$}}
\def\partt{\raise.15ex\hbox{$\widetilde$}{\kern-.37em\hbox{$\partial$}}}
\def\parts{\raise.15ex\hbox{$/$}{\kern-.6em\hbox{$\partial$}}}
\def\nablas{\raise.15ex\hbox{$/$}{\kern-.6em\hbox{$\nabla$}}}
\def\oprod{\hbox{$\rm O$}{\kern -0.8em\hbox{$\Pi$}}}
\def\partw#1{\raise.15ex\hbox{$/$}{\kern-.6em\hbox{${#1}$}}}

\def\refto#1{$^{#1}$}

\def\si{{\sigma}}

\def\gtappr{{{\lower4pt\hbox{$>$} } \atop \widetilde{ \ \ \ }}}
\def\ltappr{{{\lower4pt\hbox{$<$} } \atop \widetilde{ \ \ \ }}}

\def\topppageno1{\global\footline={\hfil}\global\headline
={\ifnum\pageno<\firstpageno{\hfil}\else{\hss\twelverm --\ \folio
\ --\hss}\fi}}

\def\toppageno2{\global\footline={\hfil}\global\headline
={\ifnum\pageno<\firstpageno{\hfil}\else{\rightline{\hfill\hfill
\twelverm \ \folio
\ \hss}}\fi}}
\def \ra{\rangle}
\def\la{\langle}

\def\dg{{^
{\dag}}}

\def\ra{\rangle}
\def\la{\langle}

\def\1{{\bf 1}}
\def\2{{\bf 2}}

\def\rarrow{\rightarrow}

\def\ul{\underline}

\def\vk{\vec k}

\def\ell{{\it l } {\rm n}}

\def\si{\sigma}

\def\cx2{\sqrt{c^2_x+c^2_y}}

\def\gkk{\gamma _{\vec k}}
\def\gk2{\gkk ^2}
\def\dw{\downarrow}
\def\up{\uparrow}
\def\gtappr{{{\lower4pt\hbox{$>$} } \atop \widetilde{ \ \ \ }}}
\def\ltappr{{{\lower4pt\hbox{$<$} } \atop \widetilde{ \ \ \ }}}

\def\pbar{{\partial\kern-1.2ex\raise0.25ex\hbox{/}}}

\def\up{\uparrow}
\def\dw{\downarrow}

\def\dg{{^{\dag}}}

\def\ra{\rangle}
\def\la{\langle}

\def\1{{\bf 1}}
\def\2{{\bf 2}}

\def\rarrow{\rightarrow}

\def\ul{\underline}

\def\vk{\vec k}

\def\ell{{\it l } {\rm n}}

\def\si{\sigma}

\def\cx2{\sqrt{c^2_x+c^2_y}}

\def\gkk{\gamma _{\vec k}}
\def\gk2{\gkk ^2}
\def\gtappr{{{\lower4pt\hbox{$>$} } \atop \widetilde{ \ \ \ }}}
\def\ltappr{{{\lower4pt\hbox{$<$} } \atop \widetilde{ \ \ \ }}}

\catcode`@=11
\newcount\tagnumber\tagnumber=0

\immediate\newwrite\eqnfile
\newif\if@qnfile\@qnfilefalse
\def\write@qn#1{}
\def\writenew@qn#1{}
\def\w@rnwrite#1{\write@qn{#1}\message{#1}}
\def\@rrwrite#1{\write@qn{#1}\errmessage{#1}}

\def\taghead#1{\gdef\t@ghead{#1}\global\tagnumber=0}
\def\t@ghead{}

\expandafter\def\csname @qnnum-3\endcsname
  {{\t@ghead\advance\tagnumber by -3\relax\number\tagnumber}}
\expandafter\def\csname @qnnum-2\endcsname
  {{\t@ghead\advance\tagnumber by -2\relax\number\tagnumber}}
\expandafter\def\csname @qnnum-1\endcsname
  {{\t@ghead\advance\tagnumber by -1\relax\number\tagnumber}}
\expandafter\def\csname @qnnum0\endcsname
  {\t@ghead\number\tagnumber}
\expandafter\def\csname @qnnum+1\endcsname
  {{\t@ghead\advance\tagnumber by 1\relax\number\tagnumber}}
\expandafter\def\csname @qnnum+2\endcsname
  {{\t@ghead\advance\tagnumber by 2\relax\number\tagnumber}}
\expandafter\def\csname @qnnum+3\endcsname
  {{\t@ghead\advance\tagnumber by 3\relax\number\tagnumber}}

\def\equationfile{%
  \@qnfiletrue\immediate\openout\eqnfile=\jobname.eqn%
  \def\write@qn##1{\if@qnfile\immediate\write\eqnfile{##1}\fi}
  \def\writenew@qn##1{\if@qnfile\immediate\write\eqnfile
    {\noexpand\tag{##1} = (\t@ghead\number\tagnumber)}\fi}
}

\def\callall#1{\xdef#1##1{#1{\noexpand\call{##1}}}}
\def\call#1{\each@rg\callr@nge{#1}}

\def\each@rg#1#2{{\let\thecsname=#1\expandafter\first@rg#2,\end,}}
\def\first@rg#1,{\thecsname{#1}\apply@rg}
\def\apply@rg#1,{\ifx\end#1\let\next=\relax%
\else,\thecsname{#1}\let\next=\apply@rg\fi\next}

\def\callr@nge#1{\calldor@nge#1-\end-}
\def\callr@ngeat#1\end-{#1}
\def\calldor@nge#1-#2-{\ifx\end#2\@qneatspace#1 %
  \else\calll@@p{#1}{#2}\callr@ngeat\fi}
\def\calll@@p#1#2{\ifnum#1>#2{\@rrwrite{Equation range #1-#2\space is bad.}
\errhelp{If you call a series of equations by the notation M-N, then M and
N must be integers, and N must be greater than or equal to M.}}\else%
 {\count0=#1\count1=#2\advance\count1
by1\relax\expandafter\@qncall\the\count0,%
  \loop\advance\count0 by1\relax%
    \ifnum\count0<\count1,\expandafter\@qncall\the\count0,%
  \repeat}\fi}

\def\@qneatspace#1#2 {\@qncall#1#2,}
\def\@qncall#1,{\ifunc@lled{#1}{\def\next{#1}\ifx\next\empty\else
  \w@rnwrite{Equation number \noexpand\(>>#1<<) has not been defined yet.}
  >>#1<<\fi}\else\csname @qnnum#1\endcsname\fi}

\let\eqnono=\eqno
\def\eqno(#1){\tag#1}
\def\tag#1$${\eqnono(\displayt@g#1 )$$}

\def\aligntag#1\endaligntag
  $${\gdef\tag##1\\{&(##1 )\cr}\eqalignno{#1\\}$$
  \gdef\tag##1$${\eqnono(\displayt@g##1 )$$}}

\def\eqalignno#1{\displ@y \tabskip\centering
  \halign to\displaywidth{\hfil$\displaystyle{##}$\tabskip\z@skip
    &$\displaystyle{{}##}$\hfil\tabskip\centering
    &\llap{$\displayt@gpar##$}\tabskip\z@skip\crcr
    #1\crcr}}

\def\displayt@gpar(#1){(\displayt@g#1 )}

\def\displayt@g#1 {\rm\ifunc@lled{#1}\global\advance\tagnumber by1
        {\def\next{#1}\ifx\next\empty\else\expandafter
        \xdef\csname @qnnum#1\endcsname{\t@ghead\number\tagnumber}\fi}%
  \writenew@qn{#1}\t@ghead\number\tagnumber\else
        {\edef\next{\t@ghead\number\tagnumber}%
        \expandafter\ifx\csname @qnnum#1\endcsname\next\else
        \w@rnwrite{Equation \noexpand\tag{#1} is a duplicate number.}\fi}%
  \csname @qnnum#1\endcsname\fi}

\def\ifunc@lled#1{\expandafter\ifx\csname @qnnum#1\endcsname\relax}

\let\@qnend=\end\gdef\end{\if@qnfile
\immediate\write16{Equation numbers written on []\jobname.EQN.}\fi\@qnend}

\catcode`@=12

\catcode`@=11
\newcount\r@fcount \r@fcount=0
\newcount\r@fcurr
\immediate\newwrite\reffile
\newif\ifr@ffile\r@ffilefalse
\def\w@rnwrite#1{\ifr@ffile\immediate\write\reffile{#1}\fi\message{#1}}

\def\writer@f#1>>{}
\def\referencefile{
  \r@ffiletrue\immediate\openout\reffile=\jobname.ref%
  \def\writer@f##1>>{\ifr@ffile\immediate\write\reffile%
    {\noexpand\refis{##1} = \csname r@fnum##1\endcsname = %
     \expandafter\expandafter\expandafter\strip@t\expandafter%
     \meaning\csname r@ftext\csname r@fnum##1\endcsname\endcsname}\fi}%
  \def\strip@t##1>>{}}

\def\citeall#1{\xdef#1##1{#1{\noexpand\cite{##1}}}}
\def\cite#1{\each@rg\citer@nge{#1}}	

\def\each@rg#1#2{{\let\thecsname=#1\expandafter\first@rg#2,\end,}}
\def\first@rg#1,{\thecsname{#1}\apply@rg}	
\def\apply@rg#1,{\ifx\end#1\let\next=\relax
\else,\thecsname{#1}\let\next=\apply@rg\fi\next}

\def\citer@nge#1{\citedor@nge#1-\end-}	
\def\citer@ngeat#1\end-{#1}
\def\citedor@nge#1-#2-{\ifx\end#2\r@featspace#1 
  \else\citel@@p{#1}{#2}\citer@ngeat\fi}	
\def\citel@@p#1#2{\ifnum#1>#2{\errmessage{Reference range #1-#2\space is bad.}%
    \errhelp{If you cite a series of references by the notation M-N, then M and
    N must be integers, and N must be greater than or equal to M.}}\else%
 {\count0=#1\count1=#2\advance\count1
by1\relax\expandafter\r@fcite\the\count0,%
  \loop\advance\count0 by1\relax
    \ifnum\count0<\count1,\expandafter\r@fcite\the\count0,%
  \repeat}\fi}

\def\r@featspace#1#2 {\r@fcite#1#2,}	
\def\r@fcite#1,{\ifuncit@d{#1}
    \newr@f{#1}%
    \expandafter\gdef\csname r@ftext\number\r@fcount\endcsname%
                     {\message{Reference #1 to be supplied.}%
                      \writer@f#1>>#1 to be supplied.\par}%
 \fi%
 \csname r@fnum#1\endcsname}
\def\ifuncit@d#1{\expandafter\ifx\csname r@fnum#1\endcsname\relax}%
\def\newr@f#1{\global\advance\r@fcount by1%
    \expandafter\xdef\csname r@fnum#1\endcsname{\number\r@fcount}}

\let\r@fis=\refis			
\def\refis#1#2#3\par{\ifuncit@d{#1}
   \newr@f{#1}%
   \w@rnwrite{Reference #1=\number\r@fcount\space is not cited up to now.}\fi%
  \expandafter\gdef\csname r@ftext\csname r@fnum#1\endcsname\endcsname%
  {\writer@f#1>>#2#3\par}}

\def\ignoreuncited{
   \def\refis##1##2##3\par{\ifuncit@d{##1}%
     \else\expandafter\gdef\csname r@ftext\csname
r@fnum##1\endcsname\endcsname%
     {\writer@f##1>>##2##3\par}\fi}}

\def\r@ferr{\endreferences\errmessage{I was expecting to see
\noexpand\endreferences before now;  I have inserted it here.}}
\let\r@ferences=\references
\def\references{\r@ferences\def\endmode{\r@ferr\par\endgroup}}

\let\endr@ferences=\endreferences
\def\endreferences{\r@fcurr=0
  {\loop\ifnum\r@fcurr<\r@fcount
    \advance\r@fcurr by1\relax\expandafter\r@fis\expandafter{\number\r@fcurr}%
    \csname r@ftext\number\r@fcurr\endcsname%
  \repeat}\gdef\r@ferr{}\endr@ferences}


\let\r@fend=\endpaper\gdef\endpaper{\ifr@ffile
\immediate\write16{Cross References written on []\jobname.REF.}\fi\r@fend}

\catcode`@=12

\def\reftorange#1#2#3{$^{\cite{#1}-\setbox0=\hbox{\cite{#2}}\cite{#3}}$}

\citeall\refto		
\citeall\ref		%
\citeall\Ref		%
\ignoreuncited
\tolerance=5000

\endtopmatter
\vskip 0.5truein
\title Instabilities of the Abrikosov-Suhl resonance
\vskip 0.5truein
\author P. Coleman$^1$, E. Miranda$^1$ and A. Tsvelik$^2$
\vskip 0.2truein
\affil $^1$ Serin Physics Laboratory
Rutgers University
PO Box 849
Piscataway, NJ 08855 0849
USA
\vskip 0.2truein
\affil $^2$ Department of Physics
Oxford University
Keble Road
Oxford OX1 3NP
UK
\vskip 0.2truein
\abstract{We consider the possibility that instabilities
of the Abrikosov-Suhl resonance lead to new
fixed point behavior of the Kondo effect in a lattice environment.
In one scenario, a pairing
component to the resonant scattering develops in the Kondo lattice, leading
to an odd frequency superconductor. We discuss experiments that
can discriminate between this picture and d-wave pairing,
and its relationship
to the non Fermi liquid fixed point of the overscreened Kondo model.}
\vskip 0.5truein
Paper presented at the SCES conference, San Diego, 1993 in celebration
of Harry Suhl's 70th Birthday.
\endtopmatter
\body
Harry Suhl's birthday provides
us with a welcome  opportunity to pause and look back at
early work on the Kondo effect in which he played a vital role.
The Kondo or ``s-d'' model for magnetic ions in a metallic host
dates
back to work by Zener\refto{zener,vonsovskii}. In 1956,
Kasuya\refto{kasuya} first cast the model
into its modern second quantized form
$$
H= H_{c} + J\sum_j
\vec \si_j\cdot \vec S_j \qquad\qquad(\vec \si_j\equiv
\psi\dg_{j\alpha} \vec \si_{\alpha\beta} \psi_{j \beta})
\eqno(kl)
$$
Here $H_c = \sum \eps_{\vk} \psi\dg_{\vk\si}\psi_{\vk\si}$
describes the conduction band and the exchange interaction
is written in a tight binding form.
In the early sixties  Kondo,\refto{kondo} building on
Anderson's superexchange concept\refto{anda, degennes}, studied
the properties of the model with antiferromagnetic
rather than the ferromagnetic interactions
envisaged by Zener. The famous logarithmic correction to
the electron scattering rate that he derived $$ {1 \over
\tau} = n_i \pi J^2 \rho \ \biggl[ 1 + 2  (J \rho)\ l
 {\rm n} \bigl({D\over 2 \pi T}	\bigr) \biggr]
\eqno(1)
$$
where $\rho$ is the conduction electron density of states and $n_i$ the
impurity concentration, beautifully explained the resistance minimum of
dilute
magnetic alloys and  definitively confirmed the
indirect antiferromagnetic exchange between magnetic ions
and conduction electrons.

Suhl was one of the first to appreciate the many body
implications of this logarithmic term. In the first of two
papers published in 1965\refto{suhl}
he remarked
``A divergence of this sort calls into question the stability
of the Fermi surface''.  Nowadays, we view the Kondo model
in the language of scaling theory, and   the
logarithmic terms that prompted Suhl's remark are taken
as an indication that the {\sl original} conduction
sea with unquenched  local moments is an unstable
{\sl fixed point}. Perturbation theory tells
us   that
spin fluctuations are ``anti-screening''  causing
a flow of the superexchange to strong coupling
as a function of energy,
according to the perturbative beta function
$$
{\partial J\rho \over \partial ln \Lambda} =
\beta(J\rho),\qquad\qquad  \beta(x)=  - 2 x^2 + 2
x^3+\dots\qquad(x<<1) \eqno(scale)
$$
Suhl's early work on the impurity model showed that the
logarithmic growth of the electron scattering amplitude
leads ultimately to the  development of an elastic
resonant scattering center at the Fermi energy.
The
 ``Abrikosov-Suhl'' resonance that he
predicted\reftorange{suhl}{nagaoka}{abrikosov} is now
understood to be a renormalized Friedel-Anderson
resonance.

Suhl's remark acquires a renewed significance in the light of
the discovery of heavy fermion compounds.
{\sl A priori}, the
presence of the Kondo logarithms in weak coupling tells us {\sl nothing}
about how more general Kondo models flow to strong
coupling: this depends on the topology of the
scaling flows.
Experimentally, most heavy fermion compounds
do indeed show features that demonstrate, that to some
degree or another, their behavior is dominated by a flow to a
Fermi liquid fixed point. However, almost all of
them show magnetic or superconducting instabilities at temperatures that are
substantial fractions of the Kondo temperature:
despite the itinerant aspects to these phase transitions,
they are in essence {\sl spin ordering
processes}, and require descriptions where the correlation
of the local moments with the conduction electrons is an
integral part of the ordering process.

These considerations motivate us to consider possible instabilities of
the Abrikosov Suhl (AS) resonance
in more general Kondo models where the appearance of new
relevant variables in the Hamiltonian
{\sl divert} the screening
process into  {\sl new basins of
attraction}.
A simple example of this phenomenon is
the impurity model
with an additional
screening channel coupled to the local moment,
$$ H =
\sum_{\alpha=1,2} H_{band}(\alpha)   + J\biggl[
\vec \si_1 + \lambda \vec \si_2
\biggr]
\cdot \vec
S_d \eqno(two) $$
Here $\alpha$ labels the screening channel.
A separatrix at $\lambda=1$ divides the Hamiltonian flows into two
distinct Fermi liquid basins of attraction where the AS resonance resides
in one channel,  but is absent from the other.\refto{noz, ingersent}
At $\lambda=1$ the AS resonance becomes unstable,
and the model scales to a non Fermi liquid (NFL)
quantum critical point, developing a unique  localized
real fermion mode with a fractional entropy.\refto{emery2} (Fig. 1a.).
There has been much recent interest in this fixed point in
connection with the possibility of a quadrupolar Kondo
effect in heavy fermion systems.\refto{Cox}

Does the lattice play a similarly
relevant role in the Kondo scaling process?
In the lattice, crossing a  separatrix between two basins
of attraction would imply a real phase
transition(Fig. 1b.) associated directly with
the Kondo effect.
One interesting possibility is
an instability in the Abrikosov-Suhl resonance that leads it
to develop an anomalous  scattering component in the
triplet channel, producing an odd frequency pairing component
in the electron self-energies.
$$
\ul{\Delta}(\kappa)
= i \si_2
\ul{ d}_c \left({V^2 \over 2 \omega}\right)\qquad\qquad(\ul{d}_c= [\hat
d_1
+ i \hat d_2]\cdot\vec \si)
\eqno(crazy)
$$
This  type  of ``odd frequency'' triplet pairing
was envisaged
by Berezinskii\refto{berezinskii, abrahams}.
Calculations that we now outline\refto{previous}
indicate that such a state is stable,
sharing features in common with the fixed point of the overscreened Kondo
model,
notably the development of neutral fermionic modes
at the Fermi energy that decouple from the spin and charge degrees
of freedom.
The state that is formed retains certain features reminiscent of
a superconductor containing a line of gap
zeros, making it a conceivable  alternative to the d-wave
theory of heavy fermion superconductivity.

A key step in our work is to bypass the difficulties of gauge
theory approaches to the Kondo lattice model,
replacing the commonly used Abrikosov pseudo-fermion representation
of spins with  a
more ancient, Majorana representation that avoids the
constraints.
This method, in  a different guise, was used by Spencer and
Doniach\refto{drone} in their
early
work on the Kondo model under the name of ``Drone fermions'',
but has since been neglected. The
essence of the method is to generalize the fermionic properties of
Pauli matrices to a lattice. Recall that Pauli matrices
anticommute $\{\si_a, \si_b\} = 2 \delta_{ab}$
which implies they are
real or ``Majorana'' ($\vec \si \dg = \vec \si$) fermions.
Their Fermi statistics guarantee that
the ``spin'' operator
$
\vec S =- {i \over 4} \vec \si \times \vec \si
$
is a faithful representation of a spin-$1/2$.
We generalize this property to the lattice by  introducing a three component
Majorana fermion
$\vec \eta _i$ at each site i, which satisfies the algebra
$
\{\eta^a_i, \eta^b_j\} = \delta_{ij}\delta^{ab},\quad(a,b=1,2,3).
$
The corresponding spin operators are then
$$
\vec S_j = -{i \over 2} \vec \eta_j \times \vec \eta_j\eqno(spin2)
$$

In  terms of Majorana fermions,
the Kondo exchange interaction can be written as
$$
H_{int}[j] = -{2 \over J} \hat V\dg_j \hat V_j\eqno(suggestive)
$$
where
$$
\hat V_j=\left(\matrix{\hat V_{j\up} \cr \hat V_{j\dw}}\right)=-{J \over
2} [ \vec
 \si \cdot \vec \eta_j]\psi_j
\eqno(spinor2a)
$$
is a charge $e$ spinor formed between  the conduction electron and local
moment.
This form
suggests the possibility that
electrons and local moments will condense
to develop a vacuum expectation value
$$
\left(\matrix{V_{j\up} \cr  V_{j\dw}}\right)=
\left(\matrix{\la \phi\vert \hat V_{j\up}\vert \phi\ra \cr
\la\phi\vert \hat V_{j\dw}\vert \phi\ra}\right)
\eqno(basicfact2)
$$
Defects of a charge $e$ spinor  correspond to
$\pi$ changes in the order parameter phase and
carry the same flux quantum as a charge 2e scalar.
Spinor order of this type
generates a non-trivial correlation between the
spin of the local moments and the conduction electron pair
degrees of freedom
$$
\la \tau_{\a}(x)S^{\be}(x)
\ra
=g{\cal M}_{\alpha}^{\ \beta}(x)
\qquad\qquad(\alpha, \beta = 1,2,3)\eqno(op)
$$
Here
$\vec \tau(x)$ is the
conduction electron ``isospin'': $z$ components describe
the number density,
$
\tau_3= {1 \over 2}(\rho(x)-1)$,  and  transverse components describe the
pairing
$\tau_+(x) =  \psi_{\up}^{\dg}(x)\psi_{\dw}^{\dg}(x)$.
The quantity $g\sim V_j^2/J^2$
defines the magnitude of the order parameter and
$\ul{{\cal M}}$ is an orthogonal matrix whose
orientation is set by the components of $V$.

On a bipartite lattice,
the lowest energy stable mean field solution
is produced by a staggered pairing field
of the form
$$
V_j =e^{i(\vec Q \cdot \vec R_j/2)}{V\over \sqrt{2}}\ {\cal{Z}}
\eqno(stable) $$
where $\vec Q=(\pi,\pi,\pi)$ and $\cal{Z}
$
is a unit spinor.
The mean field Hamiltonian describes a mixing between
local moments and conduction electrons, where the hybridization is a spinor:
$$
H_{mft}=\sum_{\vk} \left\{\tilde{\epsilon}_{\vk}\psi\dg_{\vk}\psi_{\vk}
+ {V\over \sqrt 2}\biggl[\psi\dg_{\vk} ( \vec \si \cdot \vec \eta_{\vk})
{\cal Z}  +{\rm  H. C.} \biggr]\right\}
+N|V|^2/J
\eqno(int2bc)
$$
(Here the staggered phase has been gauge transformed to the conduction
electrons by the replacement $\epsilon_{\vk}\rarrow \tilde{\eps}_{\vk}
=\epsilon_{\vk - \vec Q/2}$).   When a  conduction electron hybridizes
with the zero energy Majorana modes, it develops a self-energy
component proportional to $1/\omega$. Since the Majorana
fermion is neutral, the scattered fermion can
emerge  as either electron, or hole (see below), thereby developing
an odd frequency pairing term in the conduction electron self energy.

Diagrammatically, this process is represented as
$$
\def\cdots{\cdot\cdot\cdot}
\eqalign{
\relbar\joinrel
\relbar\joinrel
\relbar\joinrel\triangleright\joinrel
\relbar\joinrel
\relbar\joinrel
\bullet\cdots\cdots\cdots&\cdots\cdots\cdots\bullet\joinrel
\relbar\joinrel
\relbar\joinrel
\relbar\joinrel\triangleright\joinrel
\relbar\joinrel
\relbar\joinrel
\relbar\joinrel\quad= \ {V^2 \over 2\omega}\ \ul{1}\cr
\relbar\joinrel
\relbar\joinrel
\relbar\joinrel\triangleright\joinrel
\relbar\joinrel
\relbar\joinrel
\bullet\cdots\cdots\cdots&\cdots\cdots\cdots\bullet\joinrel
\relbar\joinrel
\relbar\joinrel
\relbar\joinrel\triangleleft\joinrel
\relbar\joinrel
\relbar\joinrel
\relbar\joinrel\quad =\  {V^2 \over 2\omega}\ \ul{\tau_1}}
\eqno(props)
$$
where a dotted line indicates the intermediate resonance and
we have used a Nambu notation to denote anomalous pairing
components.
For example, if ${\cal Z} =\left(\matrix{0\cr
i\cr}\right)$
then
the down electrons experience this resonant
pairing, with self energy
$$
\ul{\Sigma}_{\dw}(\om)= {V^2 \over \omega}\ul {\cal P}\eqno(sen)
$$
where $ \ul {\cal P} = {1 \over 2}\left[\ul{ 1} + \ul{\tau_1}
\right]$
is a  projection operator that projects out a Majorana component
of the down conduction sea: the remaining half
does {\sl not} couple to the odd-frequency triplet pairing field and forms
a novel band of decoupled gapless excitations. If we decompose the
conduction electron spinor into constituent ``Majorana'' components
$$
\psi_{\vk} = {1 \over \sqrt{2}}\biggl[\psi^o_{\vk}+ i \vec \psi_{\vk} \cdot
\vec \si
\biggr]{\cal Z}
\eqno(dec)
$$
the zeroth component,
$\psi^o_{\vk} =-{i \over \sqrt 2}\bigl[ \psi_{\vk \dw}-\psi\dg _{-\vk
\dw} \bigr]$  is decoupled from the local moments.
The other
three components of the conduction bands develop a gap $\Delta_g \sim V^2 /D$,
(where
$D$ is the bandwidth) that decouples them from the gapless mode,
rendering it both neutral and spinless.
The gapless  excitations consequently have vanishing spin/charge
coherence factors at the Fermi energy which
vanish {\sl linearly} with energy
$$
\la\vk
\vert \left\{ \matrix{ \tau^3 \cr \si^z  \cr}\right\}
\vert \vk
\ra =
\om_{\vk}
 \biggl( {2\mu \over V^2+ \mu^2}\biggr)
\quad \left(\om_{\vk} << \Delta_g \right)
\eqno(coherence2.5)
$$
where $\mu$ is the chemical potential and $\om_{\vk}$ the quasiparticle
energy.
Linear coherence factors lead to power laws in the nuclear magnetic
relaxation
$$
{1 \over T_1} \propto T^3
$$
that coexist with a linear specific heat capacity. Unlike
a d-wave
superconductor with lines of zeroes,  this $T^3$ NMR relaxation rate will
persist
even when the linear specific heat is large.
In a dirty d-wave superconductor we expect a Korringa
relaxation in the superconducting state once it develops a linear
specific heat. Fig. 2. contrasts the linear specific heat and
NMR relaxation rate for the simple mean field theory outlined
here, demonstrating these features.

As in all mean field theories,  issues of stability and fluctuations
are of paramount importance. Uniform
odd-frequency states appear in general to be unstable.\refto{previous}
Here the staggered phase
stabilizes the state, producing a finite Meissner stiffness.
Since there are no awkward gauge modes,
fluctuations about the odd-$\omega$ state are
similar to zero point fluctuations in  a Heisenberg magnet.
The corresponding ``Ising limit'' where these
fluctuations vanish is produced by adding an
additional  term to the Kondo interaction that couples
the conduction electron {\sl isospin} to the local moment as follows
$$
H = H_c + J \sum_j (\vec \si _j + \lambda \vec \tau_j )\cdot \vec S_j
\eqno(bizarre)
$$
The new term polarizes
the composite order
parameter, stabilizing a phase with $\ul{\cal M}=\ul{1}$.
Decomposing the conduction electron into four Majorana components
as in \(dec), then
$$
\vec \si_j +\vec \tau_j = -{i} \vec \psi_j\times \vec \psi_j,\qquad
\qquad \vec \si_j-\vec \tau_j ={2i}\psi^o_j \vec \psi_j
\eqno(morefacts)
$$
thus in the special case $\lambda=1$, the zeroth Majorana component
{\sl explicitly decouples} from the local moments,  as in the mean field
theory.

One fascinating feature of the this  isospin Kondo model, is its
precise equivalence to the two channel Kondo model in the one impurity
limit.
In the one impurity version of this model
the
conduction sea is one dimensional and near the Fermi surface,
spin-charge decoupling means that
the isospin and
spin of the conduction electron behave as two independent spin degrees of
freedom, precisely emulating the two screening channels of the
two-channel Kondo model.\refto{tsvelik}
This isomorphism is lost for multi-site or lattice models.
In fact, NFL properties are far more stable in the isospin
model, which
manifestly preserves the decoupling of
the neutral Majorana mode in the lattice at $\lambda=1$.
Furthermore, the gap in the other three Majorana bands will preserve
the decoupling of the neutral mode
in a finite region about $\lambda=1$.
Careful quantitative calculations of the zero point
fluctuations are required to establish if the domain of
attraction of the odd-$\omega$ state extends all the way to the
$\lambda=0$ Kondo lattice (dimensionality plays an important
role here). Clearly though, these simple
considerations establish  an important
link between the existence of a lattice analog of the
two channel Kondo fixed point
and the development of odd-frequency pairing.

Part of the work was supported by NSF grant  DMR-93-12138 and the SERC, UK.
E.\nobreak\ M. is supported by a grant from CNPq, Brazil.

\references

\refis{summary}For a general review
of heavy fermion physics, see N. Grewe and F. Steglich, {\sl Handbook on the
Physics and Chemistry of Rare Earths}, eds. K. A. Gschneider and L. Eyring),
{\bf 14},  343, (1991) (Elsevier, Amsterdam).

\refis{theory}For a
review of the theory of the normal phase, see e.g. P. A. Lee, T. M. Rice, J. W.
Serene, L. J. Sham and J. W. Wilkins,
\journal Comm. Cond. Mat. Phys., 12, 99, (1986); also
P. Fulde, J. Keller and G. Zwicknagl, {\sl  Solid. State Physics} {\bf 41},
 1 (1988).

\refis{rauchs}For the most recent comparison of the magnetic properties of all
known
heavy fermion superconductors, see M. Kyogaku, Y. Kitaoka, K. Asayama, C.
Geibel, C. Schank and F. Steglich, \jpjap 61, 2660, 1992.

\refis{steglich}F. Steglich , J. Aarts. C. D. Bredl, W. Leike,
D. E. Meshida, W. Franz \& H. Sch\"afer, \prl 43, 1892, 1976.

\refis{steglich2}C. Giebel. S. Thies, D. Kacrowski, A. Mehner,
A. Granel, B. Seidel, U. Ahnheim, R. Helfrich, K. Peters,
C. Bredl and F. Steglich, \zpb 83, 305, 1991;
C. Giebel, C. Shank, S. Thies, H. Kitazawa, C. D. Bredl, A. B\" ohm,
A. Granel, R. Caspary, R. Helfrich, U. Ahlheim, G. Weber and
F. Steglich, \zpb 84, 1, 1991.

\refis{ott}K. Andres   , J. Graebner \& H. R. Ott.\prl 35, 1779, 1975.

\refis{meyer}M. Goeppart Meyer, \pr 60, 184, 1941.

\refis{mott}N. F. Mott, \phl 30, 402, 1974.

\refis{lonz}L. Taillefer and G. G. Lonzarich, \prl 60, 1570, 1988.

\refis{spring}P. H. P. Reinders, M. Springford et al, \prl 57, 1631, 1986.

\refis{blount}E. I. Blount, \prb 60, 2935, 1985.

\refis{blandin}A. Blandin \& J. Friedel, {\sl J. Phys. Radium} ,
{\bf 19} , 573 , (1958).

\refis{anda}P. W. Anderson, \pr 124, 41, 1961.

\refis{langreth}D. Langreth, \pr 150, 516, 1966.

\refis{haldane}F. D, M. Haldane \prl 40, 416 , 1978.

\refis{wilson}K. G. Wilson, \rmp 47, 773, 1976.

\refis{Martin}R. M. Martin, \prl 48, 362, 1982.

\refis{krish}H. R. Krishnamurthy, J. Wilkins and K. G. Wilson, \prb 21 , 1003,
1980.

\refis{nunes}B. Jones and C: M. Varma, \prl 58, 842, 1987;
V. L. Lib\'ero and L. N. Oliviera, \prl 65,
2042, 1990;  \prb 42, 3167, 1990.

\refis{Phil}P. W.  Anderson, \jpc 3, 2436, 1970.

\refis{NCA}Y. Kuromoto, \zpb 52, 37, 1986 ; W. E. Bickers, D. Cox \&
J. Wilkins, \prl 54, 230, 1985.

\refis{kuromoto}Y. Kuromoto, \zpb 52, 37, 1986.

\refis{wilkins} W. E. Bickers, D. Cox \&
J. Wilkins, \prl 54, 230, 1985.

\refis{nick}N. Read \&  D. M. Newns, \jpc 29, L1055, 1983 ;
N.Read, \jpc 18, 2051, 1985.

\refis{me}P. Coleman, \prb 28, 5255, 1983.

\refis{long}P. Coleman, \prb  35, 5072, 1987.

\refis{crooks}N. Nagaosa \& P. Lee, \prl  64, 2450, 1990.

\refis{Noz}P. Nozi\`eres,\jdc 37, C1-271, 1976 ;
P. Nozi\`eres and A. Blandin, \jdp 41, 193, 1980.

\refis{noz}P. Nozi\`eres and A. Blandin, \jdp 41, 193, 1980.

\refis{morel}P. Morel \& P. W. Anderson, \pr 125, 1263, 1962.

\refis{usound}D. Bishop et al, \prl 65, 1263, 1984.

\refis{broholm}C. Broholm et al., \prl 65, 2062, 1990.

\refis{anderson}P. W. Anderson, \prb 30, 1549, 1984.

\refis{varma}C. M. Varma, {\sl Comments in Solid State Physics}
{\bf 11} 221, (1985).

\refis{volovic}G. E. Volovik  \& L.P.  Gorkov, \jetl 39, 674, 1984;
\jetp 61, 843, 1984.

\refis{lev} L. P. Gorkov, {\sl Europhysics Lett.} , Nov (1991).

\refis{miyake}K. Miyake, S. Schmitt Rink and C. M. Varma, \prb 34 , 7716, 1986.

\refis{bealmonod}M. T. B\'eal Monod, C. Borbonnais \& V.J.  Emery,
\prb 34, 7716, 1986.

\refis{momentUBE13}R. H. Heffner, D. W. Cooke \& D. E. Maclaughlin,
{\sl 5th Int. Conf. Valence Fluctuations} (1988).
\refis{blount}
E. I. Blount, \prb 32, 2935 , 1985.

\refis{Upt3theory}E. I. Blount et al, \prl 64, 3074, 1990 ;
R. Joynt, {\sl   S. Sci. Technol.} 1, 210, 1988;
W. Puttika \& R. Joynt, \prb 37, 2377, 1988;
T. A. Tokuyasi et al , \prb 41, 891, 1990;
K. Machida et al, {\sl J. Phys. Soc. Japan } {\bf 58} 4116, 1989.

\refis{norman}M. R. Norman, \journal Physica, C194, 203, 1992.

\refis{machida}K. Machida and M. Ozaki, \prl 66, 3293, 1991.

\refis{us}A. P. Ramirez, P. Coleman,   P. Chandra et al, \prl 68, 2680, 1992.

\refis{colemantrans}P. Coleman, \prl 59, 1026, 1987.

\refis{Cox}D. L. Cox, \prl 59, 1240, 1987 .

\refis{cox2}C. L. Seanan, M. B. Maple et al , \prl 67 , 2892, 1991.

\refis{tsvelik}B. Andraka \& A. M. Tsvelik, \prl 67, 2886, 1991.

\refis{gan}P. Coleman and J. Gan,
 {\sl Physica B} {\bf 171}, 3 (1991);
J. Gan and P. Coleman to be published.

\refis{sms}S. D. Bader, N. E. Phillips, D. B. McWhan, \prb 7, 4686, 1973.

\refis{takab}M. Kyogaku, Y. Kitaoka, K. Asayama, T. Takabatake and
H. Fujii, \jpjap 61, 43, 1992.

\refis{takab2}T. Takabatake, M. Nagasawa, H. Fujii, G. Kido, M. Nohara,
S. Nishigori, T. Suzuki, T. Fujita, R. Helfrich, U. Ahlheim, K. Fraas, C.
Geibel, F. Steglich, \prb 45, 5740, 1992.

\refis{ybb10}K. Sugiyama, H. Fuke, K. Kindo, K. Shimota, A. Menovsky,
J. Mydosh and M. Date, \journal J. Jap Phys. Soc, 59, 3331, 1990.

\refis{earlyins}T. Kasuya, M. Kasuya, K. Takegahara, \journal J. Less Common
Met, 127, 337, 1987.

\refis{rice} T. M. Rice and K. Ueda, \prb 34, 6420, 1986; C. M. Varma,
W. Weber and L. J. Randall, \prb 33, 1015, 1986.

\refis{franse} J. J. M. Franse, K. Kadowaki, A. Menovsky, M. Van Sprang
and A. de Visser, {\sl J. Appl. Phys.} {\bf 61}, 3380, (1987).

\refis{horn}S. Horn, \physica 171, 206, 1991.

\refis{visser}A. de Visser, J. Floquet, J.J.M. Franse, P. Haen,
K. Hasselbach, A. Lacerda and L. Taillefer, \physica   171, 190, 1991.

\refis{mignod}J. Rossat-Mignod, L. P. Regnault, J. L. Jacoud, C. Vettier,
P. Lejay and J. Floquet, \jmmm 76-77, 376, 1988.

\refis{aeppli} G. Aeppli, C. Broholm, E. Bucher and D. J. Bishop,
\physica 171, 278, 1991.

\refis{kuramoto}K. Miyake and Y. Kuramoto, \physica 171, 20 , 1991.

\refis{kuramoto2}Y. Kuramoto \& T. Watanabe, \physica 148B, 80, 1987.

\refis{aepplins}G. Aeppli, E. Bucher and  T. E. Mason,{\sl Proc. National High
Magnetic Field Conference}, eds E. Manousakis, P. Schlottmann,
P. Kumar, K. Bedell and F. M. Mueller
(Addison Wesley),  175, (1991).

\refis{aeppliins}T. Mason, G. Aeppli, A. P. Ramirez, K. N. Clausen ,
C. Broholm, N. Stucheli, E. Bucher and T. T. M. Palstra , \prl 69, 490, 1992.

\refis{aps1}Y. Dlichaoch, M. A. Lopez de la Torre, P. Visani, B. W. Lee amd
M. B. Maple, \journal Bull Am. Phys Soc, 37, 60, 1992.

\refis{aps2} J. G. Luissier,\journal Bull Am. Phys Soc, 37, 739, 1992.

\refis{allen}L. Z. Liui, J. W. Allen, C. L. Seaman, M. B. Maple,
Y. Dalichaouch, J. S. Kang, M. S. Torikachvili, M. A. Lopez de la
Torre, \prl 68, 1034, 1992.

\refis{ramirez2} A. Ramirez, to be published (1992).

\refis{andrei}P. Coleman and N. Andrei, \jpc 19, 3211, 1986.

\refis{andrei2}C. Destri and N. Andrei, \prl 52, 364, 1984.

\refis{allen}J. W. Allen and R. M. Martin, \jdc 41, C5, 1980.

\refis{batlogg}J. W. Allen, R. M. Martin, B. Batlogg and P. Wachter,
{\sl Appl. Phys.} {\bf 49}, 2078, (1978).

\refis{smb6}A. Menth, E. Buehler and T. H. Geballe, \prl 22, 295, 1969.

\refis{fazekas}S. Doniach and P. Fazekas, {\sl Phil. Mag.} to be published
in Phil. Mag. (1992).

\refis{insul1}M. F. Hundley et al. \prb 42, 6842, 1990.

\refis{insul2} S. K. Malik and D. T. Adroja, \prb 43, 6295, 1991.

\refis{aliev}F. G. Aliev et al.\jmmm 76-76, 295, 1988.

\refis{lonzarich}G. G. Lonzarich, \jmmm 76-77, 1, 1988.

\refis{martinins}R. Martin and J. W. Allen, \journal J. Appl. Phys., 50, 11,
1979.

\refis{lacroix}C. Lacroix and M. Cyrot, \prb, 43, 12906, 1991.

\refis{kotliar} M. Rozenberg, X. Y. Zhang,  and G. Kotliar, to be  published
(1991)

\refis{auerbach}
A. Auerbach and K.Levin,\prl 57, 877, 1986.

\refis{millis}
A.J. Millis and P.A. Lee, \prb 35, 3394, 1986.

\refis{Crag79}
D.M. Cragg and P.Lloyd, \jpc 12, L215, 1979.

\refis{aepplins}
T. E. Mason, G. Aeppli, A. R. Ramirez, K. M. Glausen, , C. Broholm, N.
St\"ucheli, E. Bucher \& T. M. M. Pasltra, Bell Labs preprint (1992).

\refis{maple}\refis{6}


\refis{palstra}
 T.T.M. Palstra, A.A. Menovsky. J. van den Berg, A.J.
Dirkmaat, P.H. Kes, G.J. Nieuwenhuys and J.A. Mydosh, \prl 55, 2727, 1985.

\refis{maple} M.B. Maple, J.W. Chen, Y. Dalichaouch, T. Kohara, C. Rossel,
M.S. Torikachvili, M. McElfresh and J.D. Thompson, \prl 56, 185, 1986.

\refis{broholmhfafm}H. J. Kjems and C. Broholm, \jmmm 76\&77, 371, 1988.

\refis{hfmags}$URu_2Si_2$, $U(Pt_{1-x}Pd)_3$,  $U_2Zn_{17}$ and $CeB_6$
are examples of commensurate afms;
the large moment systems $Ce(Cu_{1-x}Ni_x)_2Ge_2$ and $CeGa_2$ are examples
of incommensurate order.

\refis{hfmags2}J. Rossat-Mignod, L. P. Regnault, J. L. Jacoud, C. Vettier,
P. Lejay, J. Floquet, E. Walker, D. Jaccard and A. Amato \jmmm 76\&77, 376,
1988.

\refis{pethick}C. J. Pethick  and D. Pines, \prl 57, 118, 1986.

\refis{uthbe13cv}J. S. Kim, B.  Andraka and G. Stewart, \prb 44, 6921, 1991.

\refis{upt3cv}R. A. Fisher, S. Kim, B. F. Woodford, N. E. Phillips,
L. Taillefer, K. Hasselbach, J. Flouquet, A. L. Georgi and J. L. Smith,
\prl 62, 1411, 1989.

\refis{upt3phase}S. Adenwalla, S. W. Lin, Z. Zhao, Q. Z. Ran, J. B. Ketterson,
J. A. Sauls, L. Taillefer, D. G. Hinks, M. Levy and B. K. Sarma,
\prl 65, 2298, 1990.

\refis{maechida}K . Maechida and M. Ozaki, \prl 66, 3293, 1991.

\refis{trappmann}T. Trappmann, H.V. L\"ohneysen and L. Taillefer,
\prb 43, 13714, 1991.

\refis{steglich}F. Steglich , J. Aarts. C. D. Bredl, W. Leike,
D. E. Meshida, W. Franz \& H. Sch\"afer, \prl 43, 1892, 1976.

\refis{steglich2}C. Giebel. S. Thies, D. Kacrowski, A. Mehner,
A. Granel, B. Seidel, U. Ahnheim, R. Helfrich, K. Peters,
C. Bredl and F. Steglich, \zpb 83, 305, 1991;
C. Giebel, C. Shank, S. Thies, H. Kitazawa, C. D. Bredl, A. B\" ohm,
A. Granel, R. Caspary, R. Helfrich, U. Ahlheim, G. Weber and
F. Steglich, \zpb 84, 1, 1991.

\refis{upd3al2mom}A. Krimmel, P. Fisher, B. Roessli, H. Maletta,
C. Giebel, C. Schank, A. Grauel, A. Loidl and F. Steglich,
\zpb 86, 161, 1992.

\refis{berezinskii}V. L. Berezinskii, \journal JETP Lett. , 20, 287, 1974.

\refis{balatsky}For recent interest in odd-frequency
pairing, see
E. Abrahams, A. V. Balatsky,
\prb 45, 13125, 1992;
F. Mila and E. Abrahams, \prl 67, 2379, 1991.

\refis{miranda}P. Coleman, E. Miranda and A. Tsvelik, proceedings of
SCES92 conference, Sendai, to appear in {\it Physica B}(1993); Rutgers
University
preprint, to be published.

\refis{abrik}A. A. Abrikosov, \journal Physics, 2, 5, 1965.

\refis{suhl} H. Suhl, \pr 138A, 515, 1965.

\refis{morin}P. Morin and D. Schmitt, \pl 73A, 67, 1979.

\refis{cox3}D. Cox and A. E. Ruckenstein, private communication (1993).

\refis{fisk2}Z. Fisk, G. Aeppli et al, {\sl Solid State Comm.}, to be
published (1993).

\refis{schlessinger}Z. Schlessinger, Z. Fisk, G. Aeppli et al, preprint (1993).

\refis{grunertrans}W. P. Beyerman, A. M. Awasthi, J. P. Carini and G. Gruner,
\jmmm 76\&77, 207, 1988.

\refis{luthi}B. Luthi, B. Wolf, P. Thalmeier, W. Sixl and G. Bruls, preprint,
to be published (1993).

\refis{buyers}C. Broholm, H. Lin, P. T. Mathews, T. E. Mason, W. J. L. Buyers,
M. F. Collins, A. A. Menovsky , J. A. Mydosh and J. K. Kjems, \prb 43, 12809,
1991.


\refis{felten}R. Felten, F. Steglich, et al, \epl 2, 323, 1986.

\refis{uthbe13nmr}D. MacLaughlin, Cheng Tien, W. G. Clark, M. D. Lan,
Z. Fisk, J. L. Smith and H. R. Ott, \prl 51, 1833, 1984.

\refis{schloussky}K. Maki in ``Superconductivity'', editor R. Parks (1964).

\refis{maki2}K. Maki and P. Fulde, \pr 140, A1586, 1965.

\refis{hirschfeld}P. J. Hirschfeld, P. Wolfle and D. Einzel, \prb 37, 83, 1988.

\refis{upt3nmr}K. Asayama, Y. Kitaoka and Y. Kohori, \jmmm 76-77, 449, 1988.

\refis{hayden2}P. A. Midgley, S. M. Hayden, L. Taillefer and
H. v. L\" ohneysen, \prl 70, 678, 1993.

\refis{yosida}K. Yosida and Y. Yamada, \journal Prog. Theo. Phys., 46, 244,
1970;  {\bf 53}, 1286, (1975).

\refis{multi}  J. Gan, N. Andrei and P. Coleman,
{\sl Phys. Rev. Lett.} {\bf 70}, 686, (1993).


\refis{zuber}J. M. Drouffe and J. B. Zuber, {\it Phys. Reps} {\bf 102},
1, (1983).

\refis{cege2si2}D. Jaccard, K. Behnia, R. Cibin,L. Spendeler, J. Sierro,
R. Calemzuk, C. Marcenat, L. Schmidt, J. Flouquet, J. P. Brison and
P. Lejay, {\sl
Proc. Int. Conf. on Strongly Correlated Electron Systems}, Sendai 1992,
to be published in {\sl Physica B}, 1993.

\refis{doniach}S. Doniach, {\it Valence Instabilities and Narrow Band
Phenomena}, edited by R. Parks, 34, (Plenum 1977); S. Doniach, {\it Physica
}{\bf B91}, 231 (1977).

\refis{pwa1}P.W. Anderson, \journal Mater. Res. Bull., 8, 153, 1973.

\refis{steglich}F. Steglich , J. Aarts. C. D. Bredl, W. Leike,
D. E. Meshida, W. Franz \& H. Sch\"afer, \prl 43, 1892, 1976.

\refis{steglich2}C. Giebel. S. Thies, D. Kacrowski, A. Mehner,
A. Granel, B. Seidel, U. Ahnheim, R. Helfrich, K. Peters,
C. Bredl and F. Steglich, \zpb 83, 305, 1991;
C. Giebel, C. Shank, S. Thies, H. Kitazawa, C. D. Bredl, A. B\" ohm,
A. Granel, R. Caspary, R. Helfrich, U. Ahlheim, G. Weber and
F. Steglich, \zpb 84, 1, 1991.

\refis{ott}K. Andres   , J. Graebner \& H. R. Ott.\prl 35, 1779, 1975.

\refis{meyer}M. Goeppart Meyer, \pr 60, 184, 1941.

\refis{mott}N. F. Mott, \phl 30, 402, 1974.

\refis{lonz}L. Taillefer and G. G. Lonzarich, \prl 60, 1570, 1988.

\refis{spring}P. H. P. Reinders, M. Springford et al, \prl 57, 1631, 1986.

\refis{kurom}K. Miyake and Y. Kuramoto, \physica 171, 20 , 1991

\refis{blount}E. I. Blount, \prb 60, 2935, 1985.

\refis{blandin}A. Blandin \& J. Friedel, {\sl J. Phys. Radium} ,
{\bf 19} , 573 , (1958).

\refis{zwiknagel}G. Zwicknagl, \journal Adv. Phy., 41, 203, 1992;
N. d' Abrumenil and P. Fulde,  \jmmm  47-48,  1,  1985.

\refis{anda}P. W. Anderson, \pr 124, 41, 1961.

\refis{langreth}D. Langreth, \pr 150, 516, 1966.

\refis{haldane}F. D, M. Haldane \prl 40, 416 , 1978.

\refis{wilson}K. G. Wilson, \rmp 47, 773, 1976.

\refis{Martin}R. M. Martin, \prl 48, 362, 1982.

\refis{krish}H. R. Krishnamurthy, J. Wilkins and K. G. Wilson, \prb 21 , 1003,
1980.

\refis{nunes}B. Jones and C: M. Varma, \prl 58, 842, 1987;
V. L. Lib\'ero and L. N. Oliviera, \prl 65,
2042, 1990;  \prb 42, 3167, 1990.

\refis{Phil}P. W.  Anderson, \jpc 3, 2436, 1970.

\refis{NCA}Y. Kuromoto, \zpb 52, 37, 1986 ; W. E. Bickers, D. Cox \&
J. Wilkins, \prl 54, 230, 1985.

\refis{nick}N. Read \&  D. M. Newns, \jpc 29, L1055, 1983 ;
N.Read, \jpc 18, 2051, 1985.

\refis{me}P. Coleman, \prb 28, 5255, 1983.

\refis{long}P. Coleman, \prb  35, 5072, 1987.

\refis{Noz}P. Nozi\`eres,\jdc 37, C1-271, 1976 ;
P. Nozi\`eres and A. Blandin, \jdp 41, 193, 1980.

\refis{noz}P. Nozi\`eres and A. Blandin, \jdp 41, 193, 1980.

\refis{paradox}
P.~Nozi\`eres, \journal Ann. Phys. Fr., 10, 19, 1985.

\refis{morel}P. Morel \& P. W. Anderson, \pr 125, 1263, 1962.

\refis{usound}D. Bishop et al, \prl 65, 1263, 1984.

\refis{broholm}C. Broholm et al., \prl 65, 2062, 1990.

\refis{anderson}P. W. Anderson, \prb 30, 1549, 1984.

\refis{varma}C. M. Varma, {\sl Comments in Solid State Physics}
{\bf 11} 221, 1985.

\refis{lev} L. P. Gorkov, {\sl Europhysics Lett.} , Nov (1991).

\refis{miyake}K. Miyake, S. Schmitt Rink and C. M. Varma, \prb 34 , 7716, 1986.

\refis{bealmonod}M. T. B\'eal Monod, C. Borbonnais \& V.J.  Emery,
\prb 34, 7716, 1986.

\refis{ube13}H. R. Ott, H. Rudgier,  Z. Fisk and J. L. Smith, \prl 50, 1595,
1983.

\refis{upt3}G. R. Stewart, Z. Fisk, J. O. Willis and J. L. Smith, \prl 52, 697,
1984.

\refis{uru2si2}W. Schlabitz, J. Baumann, B. Pollit, U. Rauchschwalbe,
H. M. Mayer, U. Ahlheim and C. D. Bredl, \journal Z. Phys, B62, 171, 1986;

\refis{momentUBE13}R. H. Heffner, D. W. Cooke \& D. E. Maclaughlin,
{\sl 5th Int. Conf. Valence Fluctuations} (1988).
\refis{blount}
E. I. Blount, \prb 32, 2935 , 1985.

\refis{Upt3theory}E. I. Blount et al, \prl 64, 3074, 1990 ;
R. Joynt, {\sl   S. Sci. Technol.} 1, 210, 1988 ;
W. Puttika \& R. Joynt, \prb 37, 2377, 1988 ;
T. A. Tokuyasi et al , \prb 41, 891, 1990 ;
K. Machida et al, {\sl J. Phys. Soc. Japan}{\bf 58} 4116, 1989.

\refis{us}A. P. Ramirez, P. Coleman,   P. Chandra et al, {\sl Phys. Rev.
Lett.}, {\bf 68}, 2680, 1992.

\refis{jon}P. A. Lee, T. M. Rice, J. W. Serene, L. J. Sham and J. W. Wilkins,
\journal Comm. Cond. Mat. Phys., 12, 99, 1986; see also
P. Fulde, J. Keller and G. Zwicknagl, \journal Solid. State Physics, 41, 1,
1988;
for a more general review
of heavy fermion physics, see N. Grewe and F. Steglich, {\sl Handbook on the
Physics and Chemistry of Rare Earths}, eds. K. A. Gschneider and L. Eyring),
{\bf 14},  343, 1991 (Elsevier, Amsterdam).

\refis{science} P.W. Anderson, \journal Science, 235, 1196, 1987

\refis{Cox}D. L. Cox, \prl 59, 1240, 1987 .

\refis{cox2}C. L. Seanan, M. B. Maple et al , \prl 67 , 2892, 1991.

\refis{tsvelik}B. Andraka \& A. M. Tsvelik, \prl 67, 2886, 1991.

\refis{gan}P. Coleman and J. Gan,
 {\sl Physica B} {\bf 171}, 3 (1991);
J. Gan and P. Coleman to be published.

\refis{sms}S. D. Bader, N. E. Phillips, D. B. McWhan, \prb 7, 4686, 1973.

\refis{takab}M. Kyogaku, Y. Kitaoka, K. Asayama, T. Takabatake and
H. Fujii, \jpjap 61, 43, 1992.

\refis{takab2}T. Takabatake, M. Nagasawa, H. Fujii, G. Kido, M. Nohara,
S. Nishigori, T. Suzuki, T. Fujita, R. Helfrich, U. Ahlheim, K. Fraas, C.
Geibel, F. Steglich, \prb 45, 5740, 1992.

\refis{ybb10}K. Sugiyama, H. Fuke, K. Kindo, K. Shimota, A. Menovsky,
J. Mydosh and M. Date, \journal J. Jap Phys. Soc, 59, 3331, 1990.

\refis{earlyins}T. Kasuya, M. Kasuya, K. Takegahara, \journal J. Less Common
Met, 127, 337, 1987.

\refis{rice} T. M. Rice and K. Ueda, \prb 34, 6420, 1986; C. M. Varma,
W. Weber and L. J. Randall, \prb 33, 1015, 1986.

\refis{franse} J. J. M. Franse, K. Kadowaki, A. Menovsky, M. Van Sprang
and A. de Visser, {\sl J. Appl. Phys.} {\bf 61}, 3380, (1987).

\refis{horn}S. Horn, \physica 171, 206, 1991.

\refis{visser}A. de Visser, J. Flouquet, J.J.M. Franse, P. Haen,
K. Hasselbach, A. Lacerda and L. Taillefer, \physica   171, 190, 1991.

\refis{mignod}J. Rossat-Mignod, L. P. Regnault, J. L. Jacoud, C. Vettier,
P. Lejay and J. Flouquet, \jmmm 76-77, 376, 1988.

\refis{aeppli} G. Aeppli, C. Broholm, E. Bucher and D. J. Bishop,
\physica 171, 278, 1991.

\refis{kuramoto}K. Miyake and Y. Kuramoto, \physica 171, 20 , 1991.

\refis{kuramoto2}Y. Kuramoto \& T. Watanabe, \physica 148B, 80, 1987.

\refis{aepplins}G. Aeppli, E. Bucher and  T. E. Mason,{\sl Proc. National High
Magnetic Field Conference}, eds E. Manousakis, P. Schlottmann,
P. Kumar, K. Bedell and F. M. Mueller
(Addison Wesley),  175, (1991).

\refis{aps1}Y. Dlichaoch, M. A. Lopez de la Torre, P. Visani, B. W. Lee amd
M. B. Maple, \journal Bull Am. Phys Soc, 37, 60, 1992.

\refis{aps2} J. G. Luissier,\journal Bull Am. Phys Soc, 37, 739, 1992.

\refis{allen}L. Z. Liui, J. W. Allen, C. L. Seaman, M. B. Maple,
Y. Dalichaouch, J. S. Kang, M. S. Torikachvili, M. A. Lopez de la
Torre, \prl 68, 1034, 1992.

\refis{ramirez2} A. Ramirez, to be published (1992).

\refis{colemantrans}P. Coleman, \prl 59, 1026, 1987.

\refis{andrei}P. Coleman and N. Andrei, \jpc 19, 3211, 1986.

\refis{andrei2}C. Destri and N. Andrei, \prl 52, 364, 1984.

\refis{allen}J. W. Allen and R. M. Martin, \jdc 41, C5, 1980.

\refis{batlogg}J. W. Allen, R. M. Martin, B. Batlogg and P. Wachter,
{\sl Appl. Phys.} {\bf 49}, 2078, (1978).

\refis{smb6}A. Menth, E. Buehler and T. H. Geballe, \prl 22, 295, 1969.

\refis{fazekas}S. Doniach and P. Fazekas, {\sl Phil. Mag.} to be published
in Phil. Mag. (1992).

\refis{insul1}M. F, Hundley, P. C. Canfield, J. D. Thompson, Z. Fisk
and J. M. Lawrence, \prb 42, 6842, 1990.

\refis{insul2} S. K. Malik and D. T. Adroja, \prb 43, 6295, 1991.

\refis{aliev}F. G. Aliev, V. V. Moschalkov,
V. V. Kozyrkov, M. K. Zalyalyutdinov,
V. V. Pryadum and R. V. Scolozdra, \jmmm 76-76, 295, 1988.

\refis{hundley}P. C. Canfield, M. F. Hundley, A. Lacerda, J. D. Thompson and
Z. Fisk, to be published (1992).

\refis{lonzarich}G. G. Lonzarich, \jmmm 76-77, 1, 1988.

\refis{martinins}R. Martin and J. W. Allen, \journal J. Appl. Phys., 50, 11,
1979.

\refis{lacroix}C. Lacroix and M. Cyrot, \prb, 43, 12906, 1991.

\refis{kotliar} M. Rozenberg, X. Y. Zhang,  and G. Kotliar, to be  published
(1991)

\refis{auerbach}
A. Auerbach and K.Levin,\prl 57, 877, 1986.

\refis{millis}
A.J. Millis and P.A. Lee, \prb 35, 3394, 1986.

\refis{Crag79}
D.M. Cragg and P.Lloyd, \jpc 12, L215, 1979.

\refis{poor}P. W. Anderson, \journal Comm. S. St. Phys., 5, 72, 1973;
\jpc 3, 2346, 1970.

\refis{yuval}Anderson P. W. \& G. Yuval, \prl 45, 370, 1969;
Anderson P. W. \& G. Yuval, \prb 1, 1522, 1970;
Anderson P. W. \& G. Yuval, \jpc 4, 607, 1971.

\refis{swolf}J. R. Schrieffer and P. Wolff, \pr 149, 491, 1966.

\refis{cschr}B. Coqblin and J. R. Schrieffer, \pr 185, 847, 1969.

\refis{rkky}M. A. Ruderman and C. Kittel, \pr 78, 275, 1950;
T. Kasuya, \journal Prog. Theo. Phys., 16, 45, 1956;
K. Yosida, \pr 106, 896, 1957.

\refis{aepplins}
T. Mason, G. Aeppli, A. P. Ramirez, K. N. Clausen, C. Broholm,
N. Stucheli, E. Bucher, T. T. M. Palstra, \prl 69, 490, 1992.

\refis{maple}\refis{6}

\refis{palstra}
 T.T.M. Palstra, A.A. Menovsky. J. van den Berg, A.J.
Dirkmaat, P.H. Kes, G.J. Nieuwenhuys and J.A. Mydosh, \prl 55, 2727, 1985.

\refis{maple} M.B. Maple, J.W. Chen, Y. Dalichaouch, T. Kohara, C. Rossel,
M.S. Torikachvili, M. McElfresh and J.D. Thompson, \prl 56, 185, 1986.

\refis{broholmhfafm}H. J. Kjems and C. Broholm, \jmmm 76\&77, 371, 1988.

\refis{hfmags}$URu_2Si_2$, $U(Pt_{1-x}Pd)_3$,  $U_2Zn_17$ and $CeB_6$
are examples of commensurate afms;
the large moment systems $Ce(Cu_{1-x}Ni_x)_2Ge_2$ and $CeGa_2$ are examples
of incommensurate order.

\refis{hfmags2}J. Rossat-Mignod, L. P. Regnault, J. L. Jacoud, C. Vettier,
P. Lejay, J. Flouquet, E. Walker, D. Jaccard and A. Amato \jmmm 76\&77, 376,
1988.

\refis{proximity}S. Han, K. W. Ng, E. L. Wolf, A. Millis. J. L Smith and
Z. Fisk, \prl 57, 238, 1986.

\refis{steglich}F. Steglich , J. Aarts. C. D. Bredl, W. Leike,
D. E. Meshida, W. Franz \& H. Sch\"afer, \prl 43, 1892, 1976.

\refis{steglicha}F. Steglich et al., \prl 43, 1892, 1976.

\refis{steglich2}C. Giebel. S. Thies, D. Kacrowski, A. Mehner,
A. Granel, B. Seidel, U. Ahnheim, R. Helfrich, K. Peters,
C. Bredl and F. Steglich, \zpb 83, 305, 1991;
C. Giebel, C. Shank, S. Thies, H. Kitazawa, C. D. Bredl, A. B\" ohm,
A. Granel, R. Caspary, R. Helfrich, U. Ahlheim , G. Weber and
F. Steglich, \zpb 84, 1, 1991.

\refis{ott}K. Andres   , J. Graebner \& H. R. Ott.\prl 35, 1779, 1975.

\refis{meyer}M. Goeppart Meyer, \pr 60, 184, 1941.

\refis{mott}N. F. Mott, \phl 30, 402, 1974.

\refis{lonz}L. Taillefer and G. G. Lonzarich, \prl 60, 1570, 1988.

\refis{spring}P. H. P. Reinders, M. Springford et al, \prl 57, 1631, 1986.

\refis{blount}E. I. Blount, \prb 60, 2935, 1985.

\refis{blandin}A. Blandin \& J. Friedel, {\sl J. Phys. Radium} ,
{\bf 19} , 573 , (1958).

\refis{anda}P. W. Anderson, \pr 124, 41, 1961.

\refis{langreth}D. Langreth, \pr 150, 516, 1966.

\refis{haldane}F. D, M. Haldane \prl 40, 416 , 1978.

\refis{wilson}K. G. Wilson, \rmp 47, 773, 1976.

\refis{Martin}R. M. Martin, \prl 48, 362, 1982.

\refis{krish}H. R. Krishnamurthy, J. Wilkins and K. G. Wilson, \prb 21 , 1003,
1980.

\refis{nunes}B. Jones and C: M. Varma, \prl 58, 842, 1987;
V. L. Lib\'ero and L. N. Oliviera, \prl 65,
2042, 1990;  \prb 42, 3167, 1990.

\refis{Phil}P. W.  Anderson, \jpc 3, 2436, 1970.

\refis{NCA}Y. Kuromoto, \zpb 52, 37, 1986 ; W. E. Bickers, D. Cox \&
J. Wilkins, \prl 54, 230, 1985.

\refis{nick}N. Read \&  D. M. Newns, \jpc 29, L1055, 1983 ;
N.Read, \jpc 18, 2051, 1985.

\refis{me}P. Coleman, \prb 28, 5255, 1983.

\refis{long}P. Coleman, \prb  35, 5072, 1987.

\refis{Noz}P. Nozi\`eres,\jdc 37, C1-271, 1976 ;
P. Nozi\`eres and A. Blandin, \jdp 41, 193, 1980.

\refis{noz}P. Nozi\`eres and A. Blandin, \jdp 41, 193, 1980.

\refis{morel}P. Morel \& P. W. Anderson, \pr 125, 1263, 1962.

\refis{usound}D. Bishop et al, \prl 65, 1263, 1984.

\refis{broholm}C. Broholm et al., \prl 65, 2062, 1990.

\refis{anderson}P. W. Anderson, \prb 30, 1549, 1984.

\refis{varma}C. M. Varma, {\sl Comments in Solid State Physics}
{\bf 11} 221, 1985.

\refis{volovic}G. E. Volovik  \& L.P.  Gorkov, \jetl 39, 674, 1984;
\jetp 61, 843, 1984.

\refis{lev} L. P. Gorkov, {\sl Europhysics Lett.} , Nov (1991).

\refis{miyake}K. Miyake, S. Schmitt Rink and C. M. Varma, \prb 34 , 7716, 1986.

\refis{bealmonod}M. T. B\'eal Monod, C. Borbonnais \& V.J.  Emery,
\prb 34, 7716, 1986.

\refis{momentUBE13}R. H. Heffner, D. W. Cooke \& D. E. Maclaughlin,
{\sl 5th Int. Conf. Valence Fluctuations} (1988).
\refis{blount}
E. I. Blount, \prb 32, 2935 , 1985.

\refis{Upt3theory}E. I. Blount et al, \prl 64, 3074, 1990 ;
R. Joynt, {\sl   S. Sci. Technol.} 1, 210, 1988 ;
W. Puttika \& R. Joynt, \prb 37, 2377, 1988 ;
T. A. Tokuyasi et al , \prb 41, 891, 1990 ;
K. Machida et al, {\sl J. Phys. Soc. Japan }{\  \bf 58} 4116, 1989.

\refis{us}A. P. Ramirez, P. Coleman,   P. Chandra et al, {\sl Phys. Rev.
Lett.}, {\bf 68}, 2680, 1992.

\refis{Cox}D. L. Cox, \prl 59, 1240, 1987 .

\refis{cox2}C. L. Seanan, M. B. Maple et al , \prl 67 , 2892, 1991.

\refis{tsvelik}B. Andraka \& A. M. Tsvelik, \prl 67, 2886, 1991.

\refis{gan}P. Coleman and J. Gan,
 {\sl Physica B} {\bf 171}, 3 (1991);
J. Gan and P. Coleman to be published.

\refis{sms}S. D. Bader, N. E. Phillips, D. B. McWhan, \prb 7, 4686, 1973.

\refis{takab}M. Kyogaku, Y. Kitaoka, K. Asayama, T. Takabatake and
H. Fujii, \jpjap 61, 43, 1992.

\refis{takab2}T. Takabatake, M. Nagasawa, H. Fujii, G. Kido, M. Nohara,
S. Nishigori, T. Suzuki, T. Fujita, R. Helfrich, U. Ahlheim, K. Fraas, C.
Geibel, F. Steglich, \prb 45, 5740, 1992.

\refis{ybb10}K. Sugiyama, H. Fuke, K. Kindo, K. Shimota, A. Menovsky,
J. Mydosh and M. Date, \journal J. Jap Phys. Soc, 59, 3331, 1990.

\refis{earlyins}T. Kasuya, M. Kasuya, K. Takegahara, \journal J. Less Common
Met, 127, 337, 1987.

\refis{rice} T. M. Rice and K. Ueda, \prb 34, 6420, 1986; C. M. Varma,
W. Weber and L. J. Randall, \prb 33, 1015, 1986.

\refis{franse} J. J. M. Franse, K. Kadowaki, A. Menovsky, M. Van Sprang
and A. de Visser, {\sl J. Appl. Phys.} {\bf 61}, 3380, (1987).

\refis{horn}S. Horn, \physica 171, 206, 1991.

\refis{visser}A. de Visser, J. Floquet, J.J.M. Franse, P. Haen,
K. Hasselbach, A. Lacerda and L. Taillefer, \physica   171, 190, 1991.

\refis{mignod}J. Rossat-Mignod, L. P. Regnault, J. L. Jacoud, C. Vettier,
P. Lejay and J. Floquet, \jmmm 76-77, 376, 1988.

\refis{aeppli} G. Aeppli, C. Broholm, E. Bucher and D. J. Bishop,
\physica 171, 278, 1991.

\refis{kuramoto}K. Miyake and Y. Kuramoto, \physica 171, 20 , 1991.

\refis{kuramoto2}Y. Kuramoto \& T. Watanabe, \physica 148B, 80, 1987.

\refis{aepplins}G. Aeppli, E. Bucher and  T. E. Mason,{\sl Proc. National High
Magnetic Field Conference}, eds E. Manousakis, P. Schlottmann,
P. Kumar, K. Bedell and F. M. Mueller
(Addison Wesley),  175, (1991).

\refis{aps1}Y. Dlichaoch, M. A. Lopez de la Torre, P. Visani, B. W. Lee amd
M. B. Maple, \journal Bull Am. Phys Soc, 37, 60, 1992.

\refis{aps2} J. G. Luissier,\journal Bull Am. Phys Soc, 37, 739, 1992.

\refis{allen}L. Z. Liui, J. W. Allen, C. L. Seaman, M. B. Maple,
Y. Dalichaouch, J. S. Kang, M. S. Torikachvili, M. A. Lopez de la
Torre, \prl 68, 1034, 1992.

\refis{ramirez2} A. Ramirez, to be published (1992).

\refis{andrei}P. Coleman and N. Andrei, \jpc 19, 3211, 1986.

\refis{andrei2}C. Destri and N. Andrei, \prl 52, 364, 1984

\refis{allen}J. W. Allen and R. M. Martin, \jdc 41, C5, 1980.

\refis{batlogg}J. W. Allen, R. M. Martin, B. Batlogg and P. Wachter,
{\sl Appl. Phys.} {\bf 49}, 2078, (1978).

\refis{smb6}A. Menth, E. Buehler and T. H. Geballe, \prl 22, 295, 1969.

\refis{fazekas}S. Doniach and P. Fazekas, {\sl Phil. Mag.} to be published
in Phil. Mag. (1992).

\refis{insul1}M. F, Hundley, P. C. Canfield, J. D. Thompson, Z. Fisk
and J. M. Lawrence, \prb 42, 6842, 1990.

\refis{insul2} S. K. Malik and D. T. Adroja, \prb 43, 6295, 1991.

\refis{aliev}F. G. Aliev, V. V. Moschalkov,
V. V. Kozyrkov, M. K. Zalyalyutdinov,
V. V. Pryadum and R. V. Scolozdra, \jmmm 76-76, 295, 1988.

\refis{lonzarich}G. G. Lonzarich, \jmmm 76-77, 1, 1988.

\refis{martinins}R. Martin and J. W. Allen, \journal J. Appl. Phys., 50, 11,
1979.

\refis{lacroix}C. Lacroix and M. Cyrot, \prb, 43, 12906, 1991.

\refis{kotliar} M. Rozenberg, X. Y. Zhang,  and G. Kotliar, to be  published
(1991)

\refis{auerbach}
A. Auerbach and K.Levin,\prl 57, 877, 1986.

\refis{millis}
A.J. Millis and P.A. Lee, \prb 35, 3394, 1986.

\refis{Crag79}
D.M. Cragg and P.Lloyd, \jpc 12, L215, 1979.

\refis{aepplins}
T. E. Mason, G. Aeppli, A. R. Ramirez, K. M. Glausen, , C. Broholm, N.
St\"ucheli, E. Bucher \& T. M. M. Pasltra, Bell Labs preprint (1992).

\refis{maple}\refis{6}

\refis{palstra}
 T.T.M. Palstra, A.A. Menovsky. J. van den Berg, A.J.
Dirkmaat, P.H. Kes, G.J. Nieuwenhuys and J.A. Mydosh, \prl 55, 2727, 1985.

\refis{maple} M.B. Maple, J.W. Chen, Y. Dalichaouch, T. Kohara, C. Rossel,
M.S. Torikachvili, M. McElfresh and J.D. Thompson, \prl 56, 185, 1986.

\refis{broholmhfafm}H. J. Kjems and C. Broholm, \jmmm 76\&77, 371, 1988.

\refis{hfmags}$URu_2Si_2$, $U(Pt_{1-x}Pd)_3$,  $U_2Zn_{17}$ and $CeB_6$
are examples of commensurate afms;
the large moment systems $Ce(Cu_{1-x}Ni_x)_2Ge_2$ and $CeGa_2$ are examples
of incommensurate order.

\refis{hfmags2}J. Rossat-Mignod, L. P. Regnault, J. L. Jacoud, C. Vettier,
P. Lejay, J. Floquet, E. Walker, D. Jaccard and A. Amato \jmmm 76\&77, 376,
1988.

\refis{emery2} V.J.  Emery and S. Kivelson, \prb 46, 10812, 1992.

\refis{previous}P. Coleman, E. Miranda and  A. Tsvelik,
{\sl Phys. Rev. Lett.}, {\bf 70}, 2960 (1993); P. Coleman, E. Miranda and  A.
Tsvelik,
submitted to {\sl Phys. Rev. B}, Sissa preprint cond-mat/9302018.

\refis{ingersent}K. Ingersent, B. A. Jones and J. W. Wilkins, \prl 69, 2594,
1992.

\refis{drone}H. J. Spencer and S. Doniach, \prl 18, 23, 1967; D. C. Mattis,
{\it Theory of Magnetism}, (New York, Harper and Row), p. 78., (1965).

\refis{zener}C. Zener, \pr 87,  440, 1951.

\refis{vonsovskii}The notion of an s-d exchange was also developed
by Vonsovskii in the Former Soviet Union, see S. V. Vonsovskii et al.,
\jetp 2, 26, 1956 and references therein.

\refis{degennes}De Gennes also considered the effects of superexchange
in the context of superconductivity, see P. G. de Gennes, \journal
J. Phys. Rad., 23, 510, 1962.

\refis{kasuya}T. Kasuya, \journal Prog. Theoret. Phys., 16, 45, 1956.

\refis{nagaoka}Y. Nagaoka, \pr 138, A1112, 1965.

\refis{suhl}H. Suhl, \pr 138, A515, 1965; \journal Physics, 2, 39, 1965.

\refis{kondo}J. Kondo, \journal Prog. Theoret. Phys., 32, 37, 1964;
\journal ibid., 28, 772, 1962.

\refis{abrikosov}A. A. Abrikosov, \journal Physics, 2, 5, 1965.

\refis{aeppli3}See e.g. G. Aeppli and Z. Fisk, \journal Com. Mod. Phys.
B,  16, 155, 1992.

\refis{tsvelik}A. M. Tsvelik, \prl 69, 2142, 1992.

\refis{emery}``Composite'' operators of this form
have been proposed as order parameters for odd frequency pairing, in
the context of the multichannel Kondo model, by V.J.  Emery and S. Kivelson,
\prb 46, 10812, 1992.

\refis{private}E. Abrahams (private communication).

\refis{xtal}Many heavy fermion compounds clearly show Schottky
anomalies in their specific heat where the entropy integral
beneath corresponds to the suppression of magnetic fluctuations
into the higher crystal field states. See for example,
F. Rietschel et al, \jmmm 76\&77, 105, 1988, R. Felten et al, \journal Eur.
Phys. Let.,
2 , 323, 1986.

\refis{abrahams}A. V. Balatsky and E. Abrahams, \prb 45, 13125, 1992;
E. Abrahams, A. V. Balatsky, J. R. Schrieffer and P. B. Allen,
\prb 47, 513, 1993, see also A. V. Balatsky and J. R. Schrieffer in
proceedings of this conference.

\refis{berezinskii}V. L. Berezinskii, \journal JETP Lett. , 20, 287, 1974.

\refis{mermin}N.D. Mermin, \journal Rev. Mod. Phys., 51, 591, 1979.

\refis{ins1}M. F, Hundley, P. C. Canfield, J. D. Thompson, Z. Fisk
and J. M. Lawrence, \prb 42, 6842, 1990.

\refis{ins12}M. F. Hundley et al, \prb 42, 6842, 1990.

\refis{ins2}F. G. Aliev, V. V. Moschalkov,
V. V. Kozyrkov, M. K. Zalyalyutdinov,
V. V. Pryadum and R. V. Scolozdra, \jmmm 76-77, 295, 1988.

\refis{ins22}F. G. Aliev et al, \jmmm 76-77, 295, 1988.

\refis{ins3}S. Doniach and P. Fazekas, {\sl Phil. Mag. }, {\bf 65B}
1171 (1992).

\refis{weyl}See e.g.
R. Brauer and H. Weyl, \journal Amer. J. Math, 57, 425, 1935.
The Majorana  Fock space is simply constructed in momentum space.
Though this space is larger than
the conventional Hilbert space of spin $1/2$ operators that
commute at different sites, it remains faithful
to the algebra by {\sl replicating} the conventional Hilbert space
$2^{n}$ times for an even number of $2n$ sites.

\refis{kuramoto}Y. Kuramoto and K. Miyake,
\journal Prog. Theo. Phys. Suppl., 108, 199, 1992.

\refis{norman}M. R. Norman, \journal Physica, C194, 203, 1992.

\refis{aeppli}T. Takabatake, M. Nagasawa, H. Fujii, G. Kido,
M. Nohara, S. Nishigori, T. Suzuki, T. Fujita, R, Helfrich, U. Ahlheim,
K. Fraas, C. Geibel, and F. Steglich, \prb, 45, 5740, 1992;
T. Mason, G. Aeppli, A. P. Ramirez, K. N. Clausen, C. Broholm,
N. Stucheli, E. Bucher, T. T. M. Palstra, \prl 69, 490, 1992.

\refis{aeppli2}T. Takabatake et al, \prb, 45, 5740, 1992;
T. Mason et al, \prl 69, 490, 1992.

\refis{history}J. L. Martin, \journal Proc. Roy. Soc., A 251, 536, 1959;
R. Casalbuoni, \journal Nuovo Cimento, 33A, 389, 1976;
F. A. Berezin \& M. S. Marinov, \journal Ann. Phys., 104, 336, 1977.

\refis{linear}U. Rauschwalbe, U. Ahlheim, C. D. Bredl, H. M. Meyer and
F. Steglich \jmmm 63\&64, 447, 1987; R. A. Fisher, S. Kim, B. F. Woodfield,
N. E. Phillips, L. Taillefer, K. Hasselbach, J. Floquet, A. L. Giorgi and
J. L. Smith, \prl 62, 1411, 1989.

\refis{linear2}U. Rauschwalbe et al,\jmmm 63\&64, 447, 1987; R. A. Fisher
et al,  \prl 62, 1411, 1989.

\endreferences

\figurecaptions

\noindent {\bf Fig. 1.} (a) Scaling behavior of the two channel
Kondo model; (b) illustration of a hypothetical new fixed point in the
scaling trajectories of the Kondo lattice model.

\noindent {\bf Fig. 2.} $C_v/T$ and NMR $1/T_1$ calculated
for a variety of chemical potential values illustrating
the failure to develop a Korringa relaxation in the presence of
severe gaplessness.

\endfigurecaptions

\endit
\end